\documentclass{ws-p8-50x6-00}

\usepackage{cite,mcite}
\usepackage{graphicx}
\usepackage{ifthen}

\def\B{\mathrm{B}}

\def\epem{\mathrm{e^{+}e^{-}}}

\def\etal{\it et al.}

\def\pb{\mathrm{pb^{-1}}}
\def\pp{\mathrm{p\bar{p}}}

\def\qbar{\mathrm{\bar{q}}}
\def\qqbar{\mathrm{q\bar{q}}}

\def\thfs {\mathrm{\theta_{12}}}
\def\thst {\mathrm{\theta_{23}}}
\def\thtf {\mathrm{\theta_{31}}}

\graphicspath{{./}}

\begin{document}

\title{Study of Rapidity Gap Events \\
                      in Hadronic Z Decays\\
                       with L3 detector}

\author{Swagato Banerjee}

\address{for the L3 collaboration, CERN.\\ 
E-mail: Swagato.Banerjee@cern.ch}

%%%%%%%%%%%%%%%%%%%%%%%%%%%%%%%%%%%%%%%%%%%%%%%%%%%%%%%%%%%%%%
% You may repeat \author \address as often as necessary      %
%%%%%%%%%%%%%%%%%%%%%%%%%%%%%%%%%%%%%%%%%%%%%%%%%%%%%%%%%%%%%%

\maketitle

\abstracts{
Rapidity gaps have been studied in  well separated 
completely symmetric 3 jet events from hadronic Z decays to
search for colour singlet exchange. 
Asymmetries in particle flow and angular separation, 
defined with respect to the inter-jet regions,
are found to be sensitive to the exchange 
of a colour singlet object instead of a colour octet gluon. 
%Comparing data from the L3 experiment with Monte Carlo predictions,
%a limit is obtained on the fraction 
%of colour singlet exchange in data.
From a comparison of the distributions observed in the
L3 experiment and Monte Carlo predictions, 
a limit is obtained on the fraction 
of colour singlet exchange in data.
}

\section{Introduction}
\subsection{Colour Singlet Exchange}

\begin{figure}[htbp]
\begin{center}
\includegraphics[width=.49\textwidth]{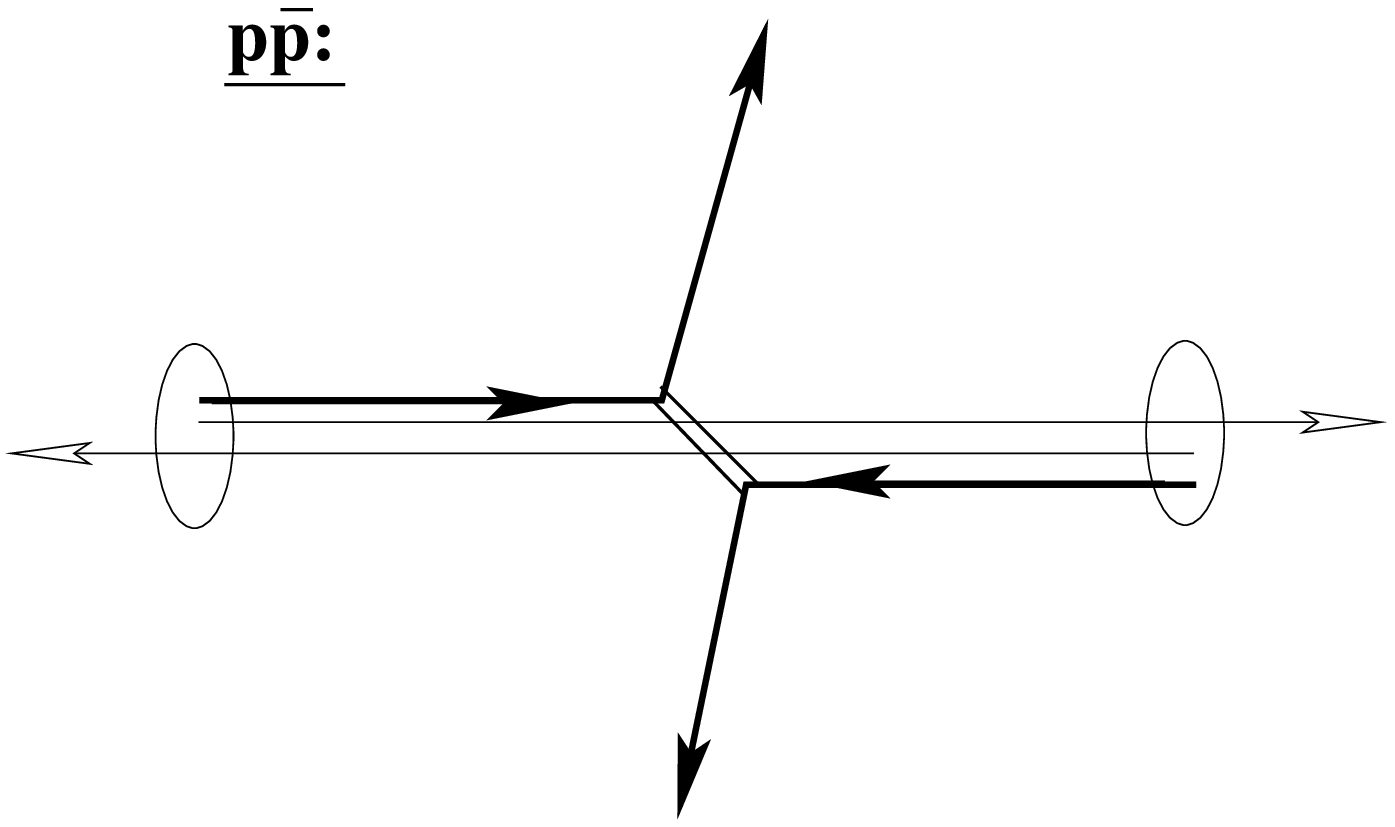}
\hfill
\includegraphics[width=.49\textwidth]{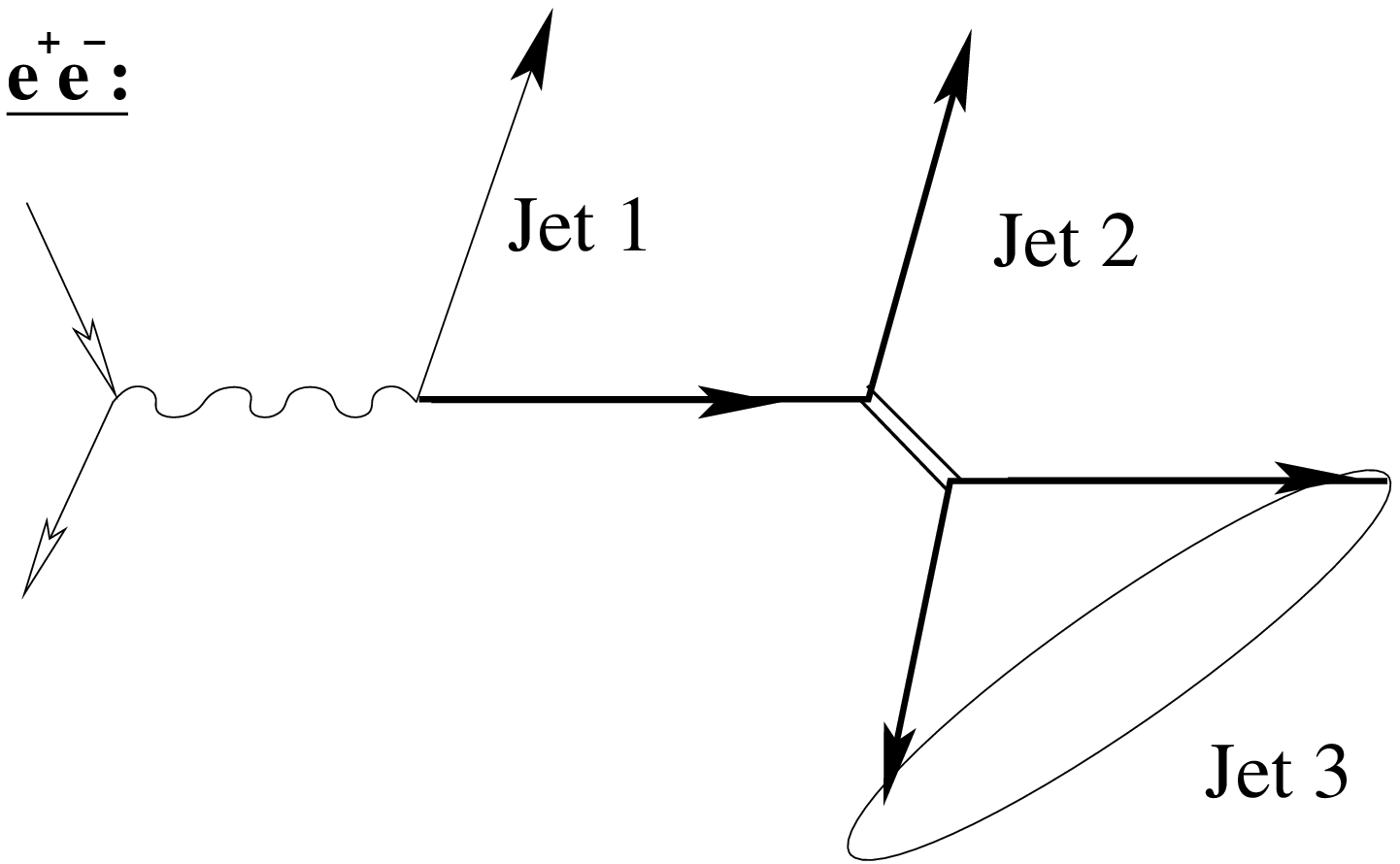}
\caption[]{{\textsc{Cse}} in $\pp$ (left) and in $\epem$ (right) reactions 
shown by double lines.}
\label{fig:cross}
\end{center}
\end{figure}

\begin{figure}[htbp]
\begin{center}
\includegraphics[width=.7\textwidth]{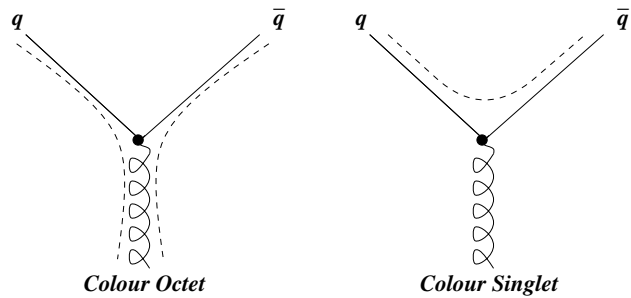}
\caption[]{Colour flow in \textsc{Coe} (left) 
 and \textsc{Cse} (right) shown by dotted lines.}
 \label{fig:cse_coe}
\end{center}
\end{figure}

Events with large rapidity gaps, attributed to 
colour singlet exchange (\textsc{Cse}),
have been observed at {\textsc{Hera}}~\cite{hera} 
and the {\textsc{Tevatron}}~\cite{tevatron}.
By crossing symmetry, we may expect similar gaps in three 
jet hadronic Z decays (figure~\ref{fig:cross}).
This study searches for such gaps by exploiting
differences in colour flow in three jet topology 
between \textsc{Cse} and colour octet exchange (\textsc{Coe}):
in the latter colour flow is present between the qg and $\qbar$g gaps and 
is inhibited by destructive interference in the $\qqbar$ gap,
while in the former case colour flow occurs only in the $\qqbar$ gap  
(figure~\ref{fig:cse_coe}).

\subsection{QCD models}
For this study, the \textsc{Jetset} Parton Shower program
%~\cite{jetset} 
has been used to model \textsc{Coe}.
Two toy models have been used to simulate the desired colour flow in 
\textsc{Cse}:
events of type $\qqbar\gamma$ with a photon effective mass as in the gluon 
jet mass distribution are generated, and the photon is then replaced 
by a boosted two jet event (model \textsc{CS0}), or 
by a gluon fragmenting independently (model \textsc{CS2}).
The \textsc{Rathsman} model, tuned to L3 hadronic Z decay data
with string re-interaction probability set to 0.1, is also 
studied~\cite{rath}.

\section{Methodology}
Three jet events are selected at fixed jet resolution
parameter, set to 0.05 in the \textsc{Jade} algorithm
%~\cite{jade} 
(default)
and to 0.01, 0.02 in the \textsc{Durham} algorithm
%~\cite{kt} 
(cross-check), with inter-jet angles within
$\pm 30^{\circ}$ from the symmetric Mercedes topology.
Particle momenta are projected onto the event plane defined
by the two most energetic jets, and then all the angles
(measured from most energetic jet) are rescaled so as to align
jets at 0$^\circ$-120$^\circ$-240$^\circ$ for
uniformity in event-to-event comparison.
To quantify the inter-jet gap angle, two definitions are used:
the minimum opening angle of the particles measured 
from the bisector in each gap ({\em{B-angle}}), 
and maximum separation angle between adjacent particles in each gap 
({\em{S-angle}}). 
%To minimise the bias from fragmentation, 
%cones of half angle = 15$^{\circ}$ (default), 20$^{\circ}$ (cross-check)
%around the jet-axis are removed in the definition of the gaps. 
Ordering the jets such that jets 1, 2 are from primary quarks, and
referring to {\em{B(S)-angle}}
between jets $i$, $j$ generically as $\theta_{ij}$ $(i,j=1,3)$,
{\em gap asymmetries} are defined as:
A$_{12}$ = $\frac{-\thfs+\thst+\thtf}{\thfs+\thst+\thtf}$,
and cyclically for the other gaps.
Reduced colour flow and thereby larger separation for \textsc{Cse} 
in gaps 23, 31 (together referred to as qg) with respect to gap 12, 
should thus make A$_{12}$ peak for positive values 
more strongly for \textsc{Cse} than for \textsc{Coe}.

\section{Results}
\begin{figure}[hbtp]
\begin{center}
\includegraphics[width=.49\textwidth]{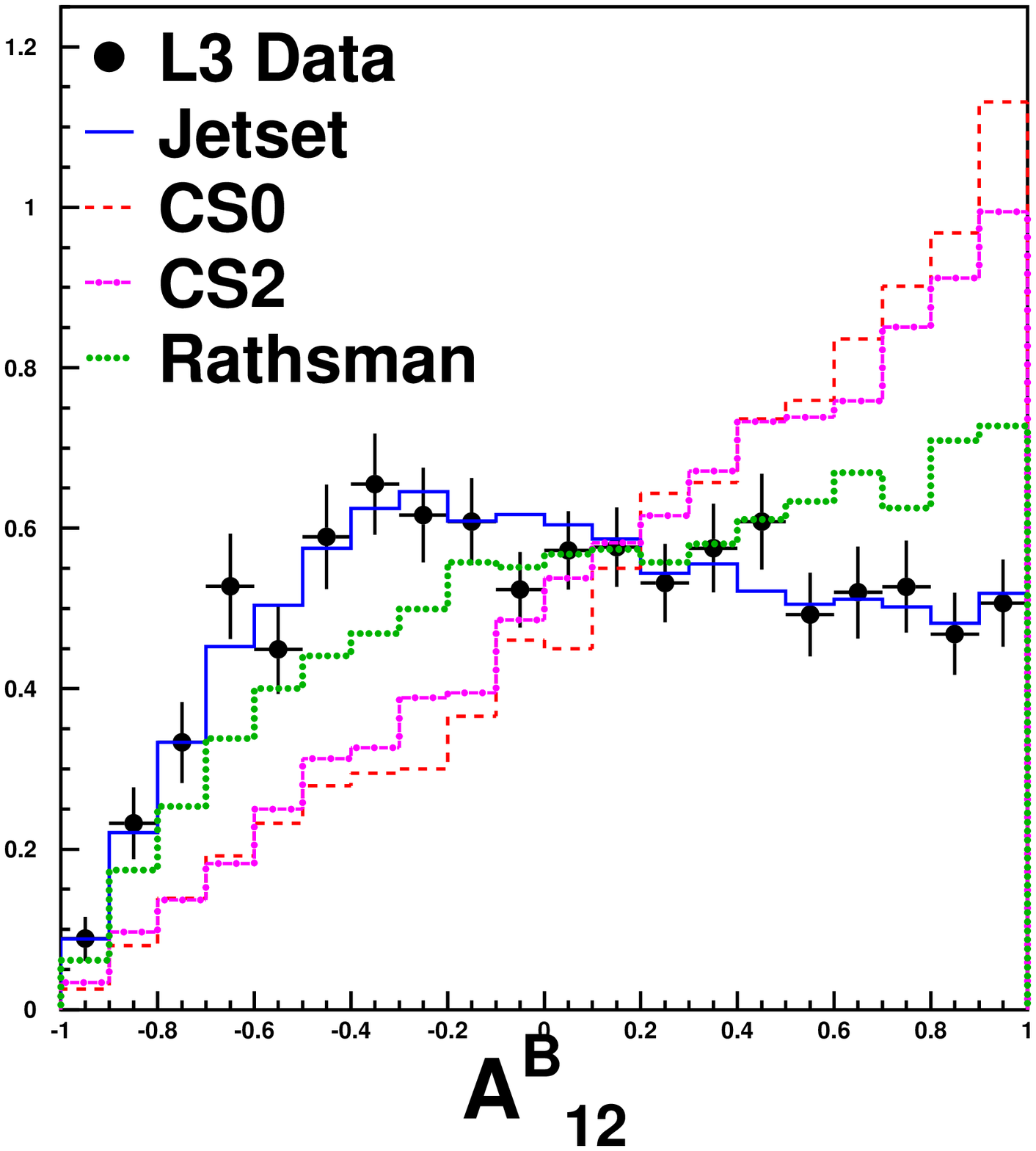}
\hfill
\includegraphics[width=.49\textwidth]{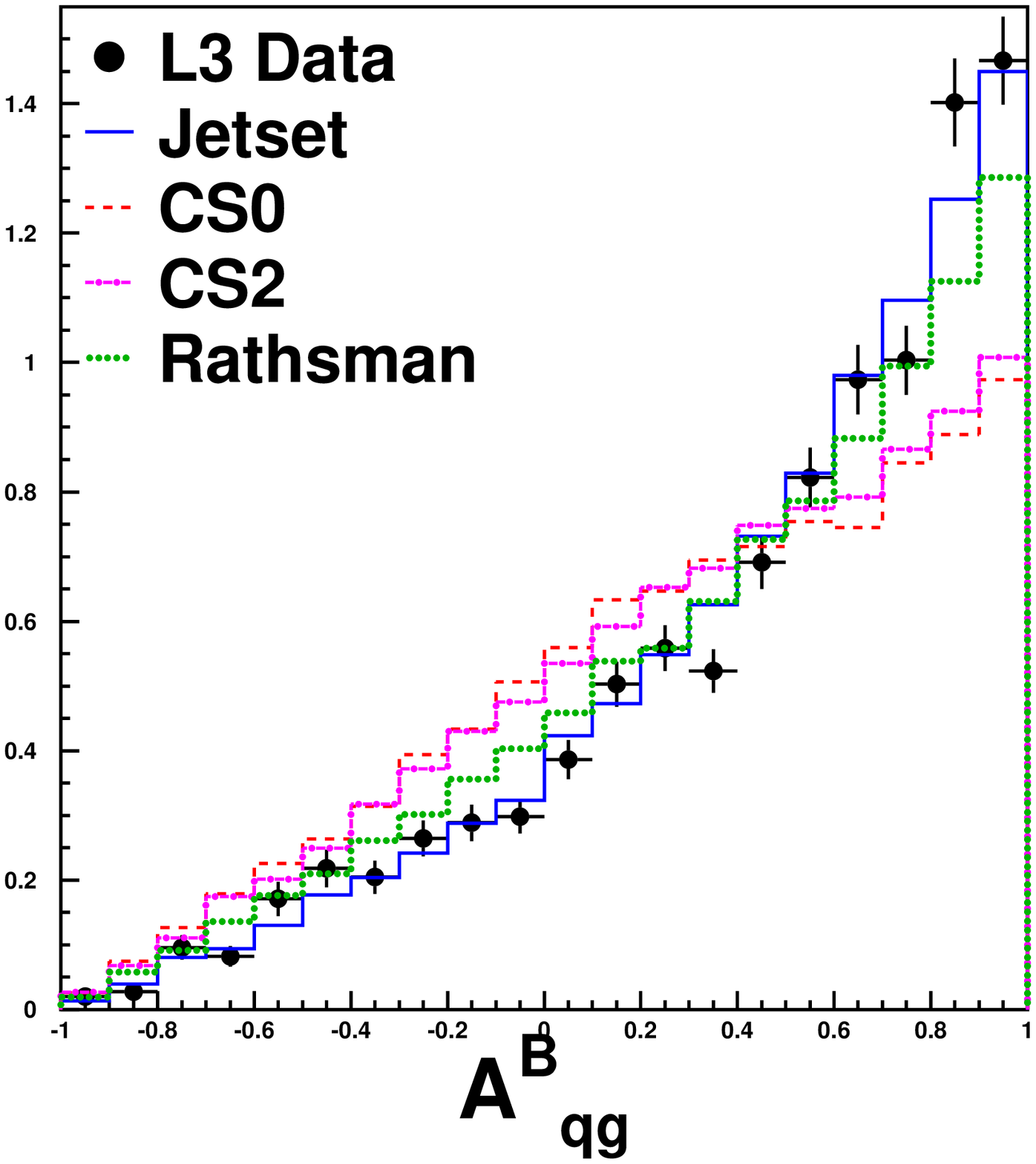}
\caption[]{Minimum bisector angle gap asymmetries for gaps 12 and qg.}
\label{fig:ab}
\end{center}
\end{figure}

\begin{figure}[hbtp]
\begin{center}
\includegraphics[width=.49\textwidth]{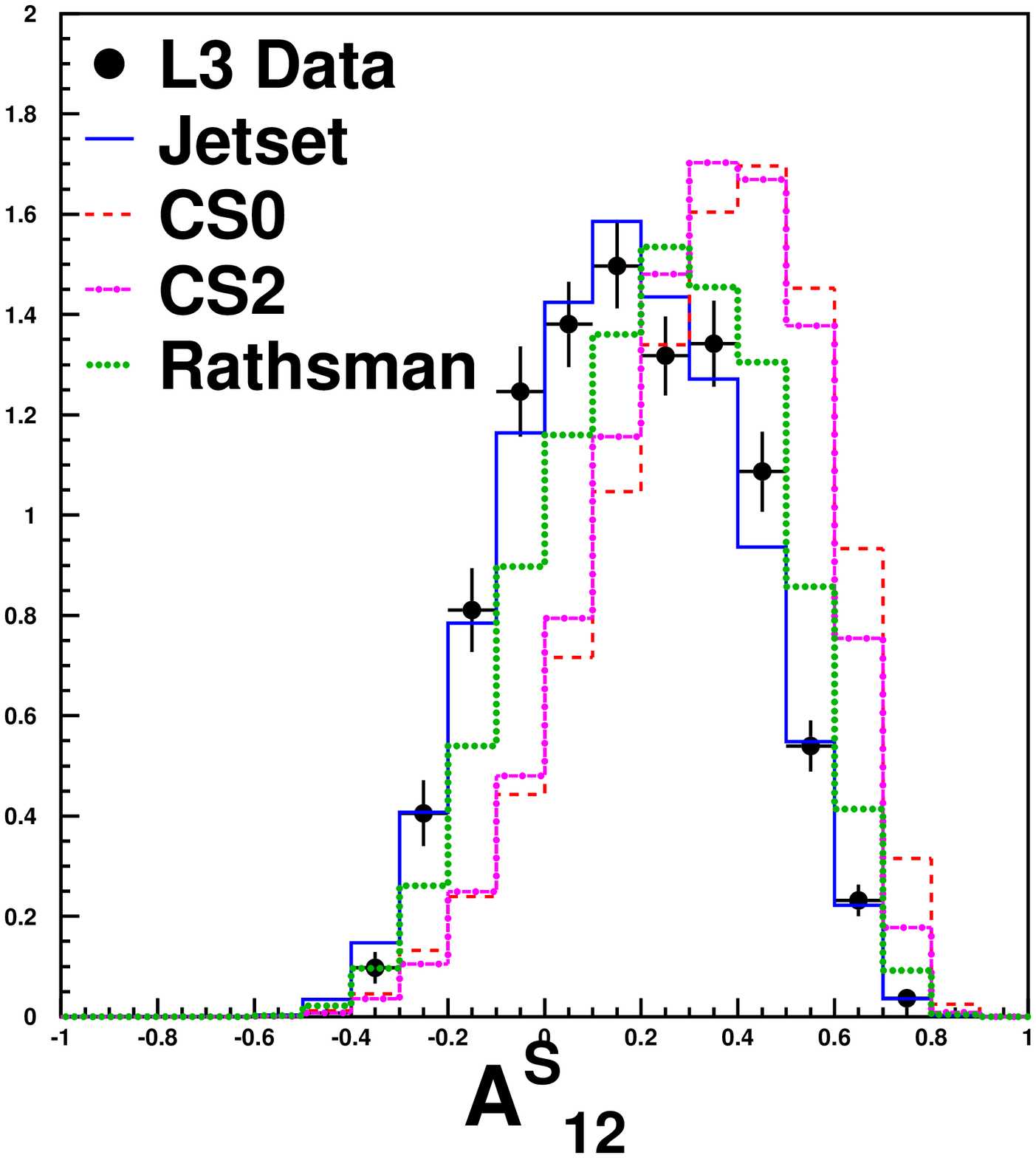}
\hfill
\includegraphics[width=.49\textwidth]{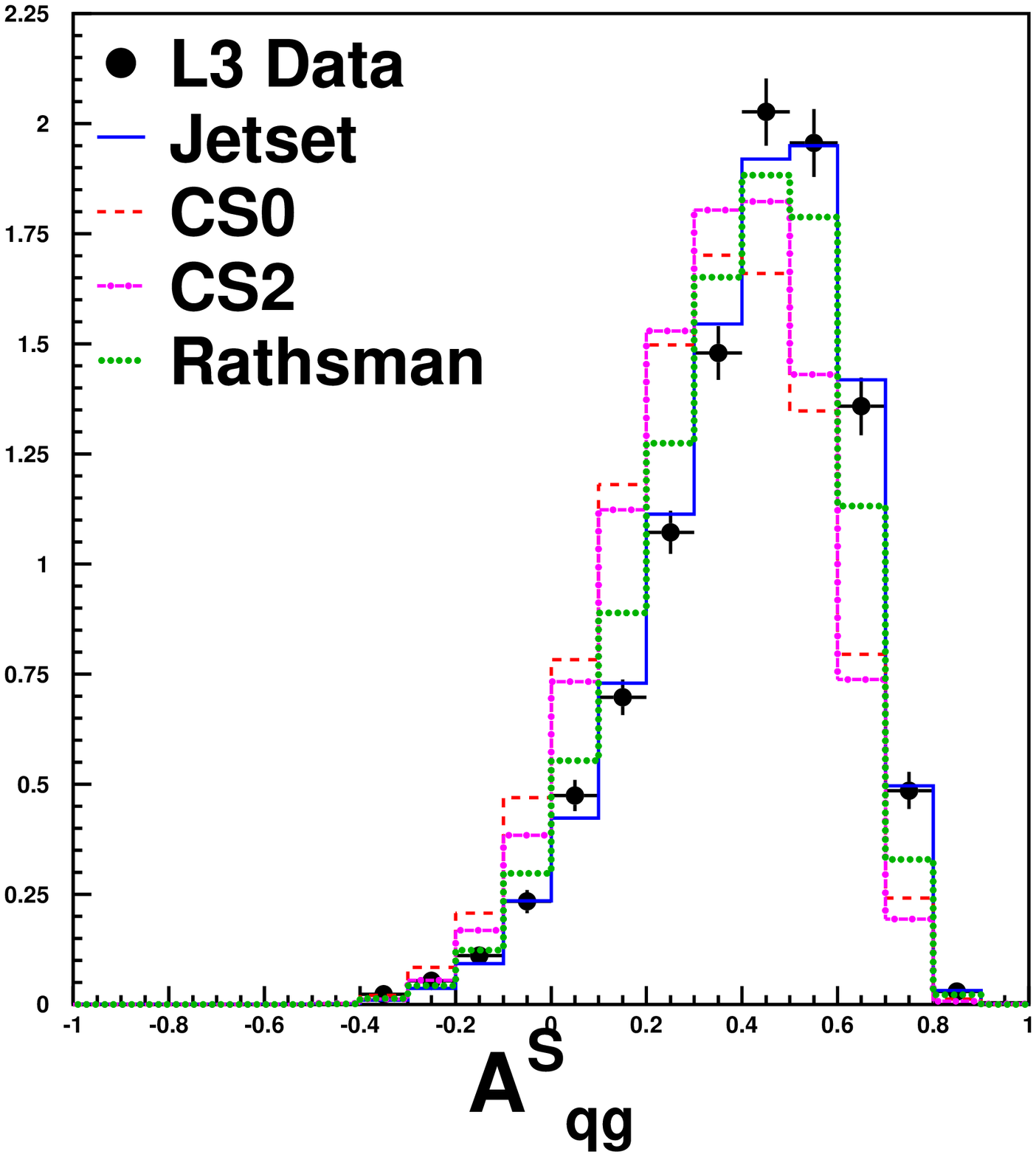}
\caption[]{Maximum separation angle gap asymmetries for gaps 12 and qg.}
\label{fig:as}
\end{center}
\end{figure}

The analysis is performed with 2 million hadronic Z decay events
recorded by the L3 detector during 1994-95 corresponding to
a luminosity of 75.1 $\pb$~\cite{l3note}.
For the three jets, labelled in order of decreasing energy,
the probability of having a gluon in jet 3
in the default jet reconstruction is estimated using \textsc{Jetset}
to be 69\%, reducing to 48\% for Mercedes topology.
In order to distinguish quark jets from gluon or colour singlet jets, 
quark jets are tagged by demanding that the
b-tag discriminants of jets 1 and 2 are above a certain cut-off ($\B_1$) 
and that of jet 3 is below a second cut-off ($\B_2$).
The selection, optimized to give the best sensitivity,
tags 2668 events with $\B_1$ = 1.25, $\B_2$ = 1.5, 
corresponding to a probability of identifying a gluon in jet 3 as 78\%, 
the probability being varied by 10\% for systematic studies.
The data asymmetry distributions, corrected for detector
effects as well as flavour composition,
using detector simulated \textsc{Jetset} events with 
quarks jets from all flavours tagged in the first two jets, 
are compared to different models,
all normalized to unit area, in figures~\ref{fig:ab},~\ref{fig:as}.
The systematic errors on bin-content from jet reconstruction ($\sim$ 4-7\%)
and tagging ($\sim$ 3-8 \%) have been added in quadrature.

\noindent 
The data are consistent with \textsc{Coe} as modelled by \textsc{Jetset}.
The probability for having a $\chi^2$ greater than the
observed value between data and \textsc{Rathsman} is
${\cal{O}}(10^{-6})$. Fits to data with admixture of 
a fraction of \textsc{Cse}
and remaining \textsc{Coe} are consistent with zero admixture.
Combining the asymmetry between gaps 12 and qg, 
the upper limit of the fraction of \textsc{Cse} present in data
is estimated
to be between 7\% to 9\% at 95\% confidence level.

\section*{Acknowledgments}
I thank Sunanda Banerjee and John Field for their 
valuable collaborative efforts in this work.

\end{document}